\documentclass[letterpaper]{article}
\usepackage{iccc}
\usepackage{graphicx}
\usepackage[ruled,vlined]{algorithm2e}
\usepackage{graphics}
\usepackage{hyperref}
\usepackage{lipsum}
\usepackage{etoolbox}
\usepackage{natbib}

\usepackage{times}
\usepackage{abstract}
\usepackage{graphicx}
\usepackage{helvet}
\usepackage{courier}
\title{Co-creation and ownership for AI radio}
\author{Skylar Gordon\textsuperscript{1}, Robert Mahari\textsuperscript{12}, Manaswi Mishra\textsuperscript{1}, Ziv Epstein\textsuperscript{1} \\ \newline \texttt{\{sfgordon, rmahari, manaswim, zive\}@mit.edu} \\
\textsuperscript{1} MIT Media Lab  \textsuperscript{2} Harvard Law School}

\begin{document} 
\maketitle
\begin{abstract}
\begin{quote}




Recent breakthroughs in AI-generated music open the door for new forms for co-creation and co-creativity. We present Artificial.fm, a proof-of-concept casual creator that blends AI-music generation, subjective ratings, and personalized recommendation for the creation and curation of AI-generated music. Listeners can rate emergent songs to steer the evolution of future music. They can also personalize their preferences to better navigate the possibility space. As a ``slow creator'' with many human stakeholders, Artificial.fm is an example of how casual creators can leverage human curation at scale to collectively navigate a possibility space. It also provides a case study to reflect on how ownership should be considered in these contexts. We report on the design and development of Artificial.fm, and provide a legal analysis on the ownership of artifacts generated on the platform. 
\end{quote}
\end{abstract}

\section{Introduction}
\begin{quote}
``As notions about the nature and function of music become embedded in the structure of software-based musical systems and compositions, interactions with these systems tend to reveal characteristics of the community of thought and culture that produced them.'' \newline
\hspace{20 mm} --George Lewis \citep{lewis2000too}
\end{quote}

Recent breakthroughs in deep learning have introduced the opportunity for generating high-fidelity songs in the raw audio domain. Some believe that this new potential portends the end of musical creativity, while others think it represents yet another tool to augment musical production. Both sides have merit, but of particular importance is the fact that these end-to-end music generation systems can synthesize music without any understanding in music composition or technique. This opens up the possibility of embedding them in computational creativity systems, which allows users to explore a large possibility space of music without formal musical training\citep{compton2015casual} and engage in \textit{mixed-initiative co-creativity}~\cite{yannakakis2014mixed}. 

To interrogate these questions, we introduce a proof-of-concept casual creator system, Artificial.fm, that allows listeners to help curate and steer the evolution of music generated with OpenAI's Jukebox model \citep{dhariwal2020jukebox}. In addition to listening to this new kind of music, listeners can also provide feedback on the generated songs, thus helping the AI learn to generate better music in the future. The system also uses these ratings to provide personalized music recommendations, which helps the music evolve to fit the preferences of the listener. 

These components form an interconnected sociotechnical system for music generation and curation, with many distinct human stakeholders. This distributed model of production complicates the definition of \textit{the user} of the system, since many different users are involved in different ways. It also raises important questions about who owns the artifacts generated by the system. 

Our system falls in the lineage of ``slow creators'' defined by a ``problematic gulf of execution'' \citep{compton2019casual}. This collection of creators involves most audio-based generators, since evaluating songs requires the user to actually listen to the outputs, instead of quickly discerning its quality, as with visuals. However, a distinct yet understudied aspect of Artificial.fm is the fact that generation itself is high-latency and therefore impossible to do on the fly: Jukebox takes about 20 hours to generate 20 seconds of audio. Thus Artificial.fm explores design patterns for a growing set of systems where intensive underlying computation means real-time interaction with the underlying generator is fundamentally infeasible. 

In this paper, we present the case study of Artificial.fm to highlight how slow creation can translate to the evaluation and curation of AI-generated music. We then use legal precedent to trace the multiple stakeholders involved in this process and unpack the each actor's stake in ownership.

\section{Related Work}

 Algorithmic Music has a rich history amongst composers starting in the pre-computing era from the process works of George Brecht's Drip Music (1962), Stockhausen’s Setz die Segel zur Sinn \citep{stockhausen} and Xenakis’ Formalised Music \citep{xenakis1992formalized} to the formation of the US League of Automatic Music Composers (1978).

 Algorithms have been used to generate music both in the symbolic domain \citep{hiller1957musical, moorer1972music, hadjeres2017deepbach,huang2018music} and in the waveform domain through digital vocoders \citep{bonada2007synthesis, blaauw2017neural} and synthesizers \citep{mehri2016samplernn, engel2017neural}.

\citet{compton2019casual} identifies Musical computational creativity systems as inherently ``slow creators'' where the user evaluation in the \textit{grokloop} is implicitly slow. Though computationally slow in generating these more complex musical generative spaces, this newer generation of  ‘slow systems’ are capable of producing more aesthetically pleasing and uniquely shaped outputs that often feel more rewarding and personal to the user. More recent casual creators like - Magenta’s Tone Transfer, Piano Genie and applications produced at the BitRate ML and Music hackathon 2020, take advantage of modern AI models to produce higher fidelity musical outputs.

These generators are characterized by large possibility spaces, which can be difficult for individuals to explore. A promising approach to rapidly search through a large possibility space to find the ``gem in the rough'' is to crowdsource its exploration. A diverse set of casual creators leverage collaborative media to produce intriguing artifacts. The Reddit R/Place experiment had users collaboratively paint a pixel canvas \citep{rappaz2018latent}. Drave's Electric Sheep used user feedback and evolutionary algorithms to generate new ``sheep'' - fractal animations adapted to crowd preferences \citep{draves2005electric}. PicBreeder also uses evolutionary algorithms and allows users to collaboratively evolve images \cite{secretan2008picbreeder, secretan2011picbreeder}. Feed the Ganimals allowed users to explore and curate AI-generated hybrid animals, and found that social cues led to the formation of diverse local trends \citep{epstein2020interpolating, epstein2021social}.

\section{System Overview}

Artificial.fm uses OpenAI’s Jukebox \citep{dhariwal2020jukebox}, a deep neural network trained on 1.2 million songs, for music generation. Jukebox has the ability to take as input a ``prime” of existing music which it then improvises on top of. We solicit primes from local musicians as part of a collaboration to support artists affected by the pandemic. Jukebox also takes in a specified artist and genre as inputs which condition the style of the generated song outputs. 

The outputs of the song generation process are streamed via the platform, where listeners provide subjective feedback on the AI-generated music, in the form of ratings. The questions related to how happy, danceable, artificial, instrumental, upbeat and song was, and how clear the lyrics, and if they liked it, on a 5-point Likert scale (see Supplementary Information Section 1.2 for more information). 

The crowdsourced feedback is then used to adapt the generation process with an algorithm that balances exploring new permutations of parameters with exploiting existing parameters that are popular with users. This is achieved using a variation of Thompson sampling, which is regret-minimizing in such contexts \citep{chapelle2011empirical}.  To do so, we use the Spotify API and Essentia \citep{bogdanov2013essentia} to generate a rich set of covariates for the artist of the prime, as well as candidate artists and genres (see Supplementary Information Section 1.3 for more information on how these covariates are generated).  



As new primes are solicited from local musicians, the following algorithm finds parameters (e.g. an artist and genre prompt) to pair with that prime to balance exploration and exploitation:  First, we fit a model $\hat{f}$ predicting ratings of the existing songs (e.g. How much do you like this song?) based on the Spotify covariates of that song's prime artist, artist prompt, and genre prompt (27 features total). Then, we sample M artist, genre pairs from the joint distribution of these prompts in the input space. Then, we predict the rating of that artist, genre pair for the given prompt $\hat{f}_{prime}(a_\ell, g_\ell)$. We then take the top $\gamma$ artist, genre pairs and randomly sample one uniformly (here $\gamma$ controls exploration vs exploitation, $\gamma=1$ is maximal exploitation, $\gamma=M$ is maximal exploration). See Supplementary Information Section 1.4 for more details about this algorithm. \footnote{This algorithm assumes there is already a large number of both songs and ratings and therefore requires solving the ``cold-start problem.'' Since the scope of this short paper is introducing the concept of AI radio via Artificial.fm with preliminary user testing and ethical considerations, this algorithm should be considered as
a sketch for how Artificial.fm would work at scale. As such, we leave formal evaluation of such an approach to music generation to future work.}

Artificial.fm also provides personalized song recommendations to users. Through a preference elicitation interface (see Figure S3), users can explicitly specify the kind of songs they would like to hear. Based on their stated preferences, a personalized recommendation algorithm serves songs to them consistent with these preferences (see Supplementary Information Section 1.5 for more details about this recommender system).


\section{Data and Results}
As of July 21, 2021, we accumulated 522 ratings of 71 songs by 40 people. The songs were generated with genre prompts from folk, house, pop, americana, rock, classical, electronic, and funk, and artist prompts from The Weeknd, Aerosmith, The Doors, Justin Bieber, Elton John, Dolly Parton, Otis Redding, and Lady Gaga. The primes were sourced from several local artists we reached out to. The 40 people found the platform through word of mouth.

\begin{figure*}[h]
    \centering
    \includegraphics[width=0.99\textwidth]{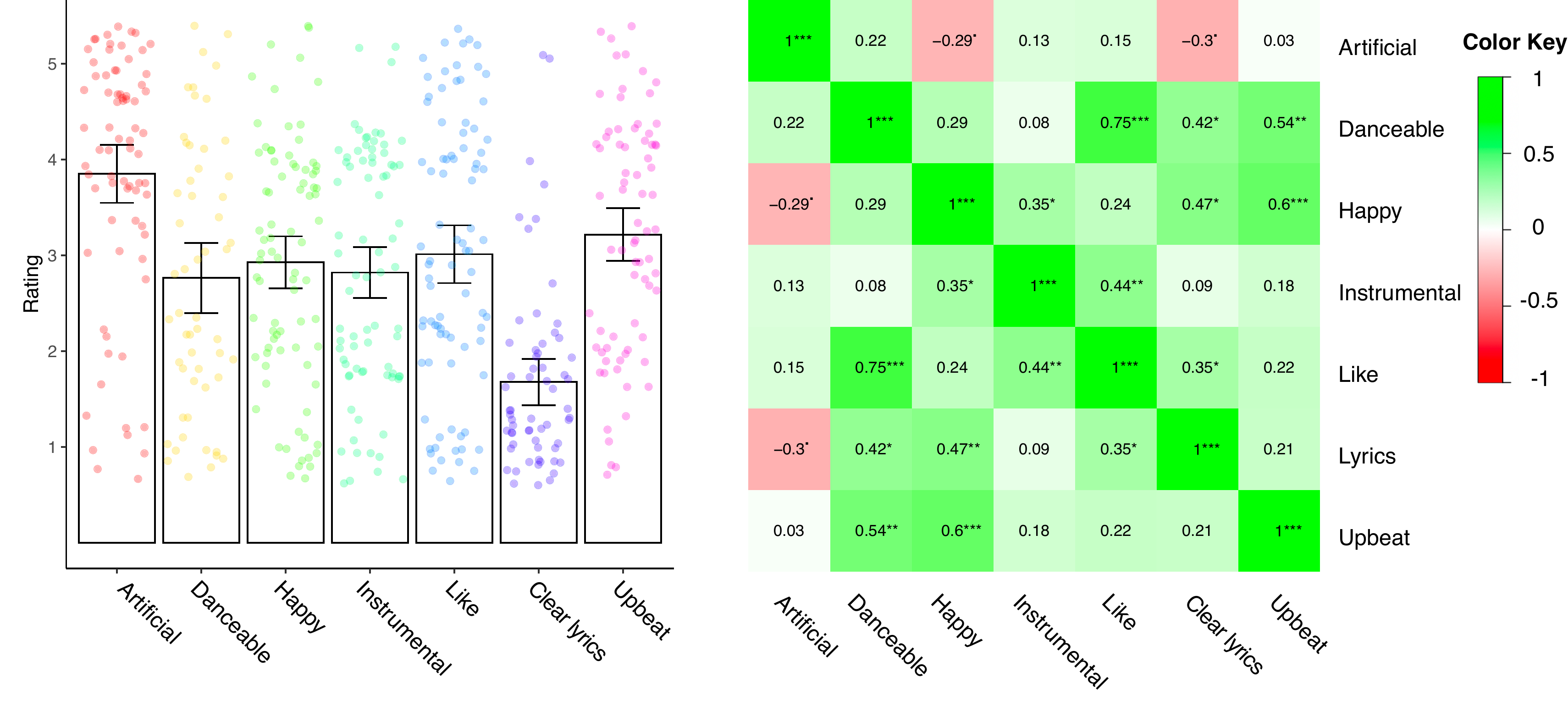}
    \caption{Left: Song ratings by question. Right: Pairwise correlation matrix between question ratings. $^{.}$ refers to $p\leq0.1$, $^*$ refers to $p\leq.05$, $^{**}$ refers to $p\leq.01$, $^{***}$ refers to $p\leq.001$.}
    \label{fig:results}
\end{figure*}

The distribution of ratings by question is shown on the left of Figure~\ref{fig:results}. Relative to the other questions, listeners found the songs highly artificial (one-sided $t$ test, $p<0.001$), and lacking in clear lyrics ($p<0.001$). This suggests that the music of Artificial.fm may not fall into the ``normal distribution'' of what you find on the radio, but instead represents a polyphonous new kind of music onto itself. That being said, we did find meaningful variation in all seven questions ratings, which suggests there is quantifiable diversity in the possibility space to explore and optimize.   

The pairwise correlations between these seven questions are shown on the right of Figure~\ref{fig:results}. We observe that perceptions of liking a song is associated with ratings of a song being danceable ($R = 0.75$, $p<0.001$), instrumental ($R = 0.44$, $p=0.004$), and having clear lyrics ($R = 0.35$, $p=0.037$). We also find that ratings of the artificiality of a song are marginally negatively associated with having clear lyrics ($R = -0.30$, $p=0.088$), and how happy the song is perceived to be ($R = -0.29$, $p=0.078$).

\section{Ownership of AI-Generated Music}
The owner of a casual creators' output should be the entity responsible for creation.
\citet{compton2015casual} defines casual creation as the ``the creation [of] new artifacts that bring feelings of pride, ownership, and creativity to the users that make them.'' This definition centers the users of a casual creator as the owners of its output. Artificial.fm challenges this idea of ownership and highlights open questions related to ownership of AI-generated works. \footnote{We use the term ``ownership" broadly to encompass all the rights commonly associated with authorship. Where relevant, this section will base its analysis on U.S. and California law.} 

At least five actors could claim some level of ownership over the works created by Artificial.fm: (1) the artist who submitted the prime on which a piece of music is based, (2) the many artists whose music was used to train Jukebox, (3) the system architects who developed Artificial.fm, (4) the listeners whose ratings are used to steer music production, and (5) the artificial intelligence itself. This section begins to explore the question of ownership for casual creators by analyzing the legal basis on which these actors may claim ownership and concludes by suggesting ownership models better suited to the distributed nature of systems like Artificial.fm.

\subsection{The Prime and Training Artists \includegraphics[width=5mm]{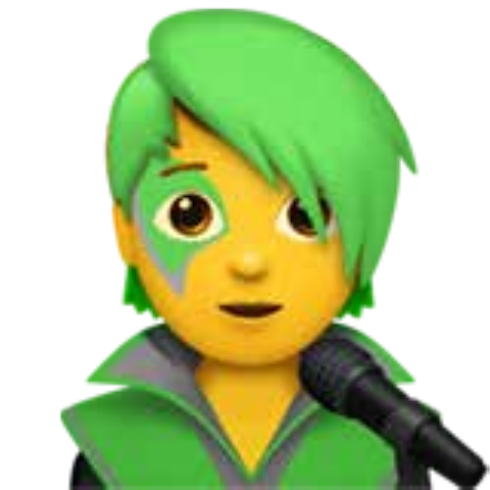}\includegraphics[width=5mm]{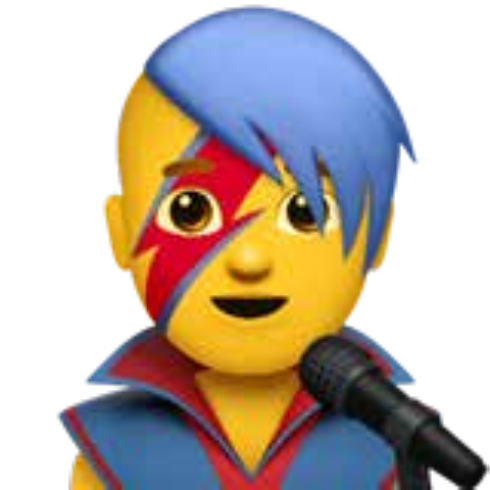}}
Both the prime and training data artists could claim ownership over a given piece of music created by Artificial.fm by arguing that Artificial.fm infringes on their copyright. To this end, they would need to show that the generated music is ``substantially similar" to their work \citep{williamsvgaye} and that the music was not independently created \citep{FeistvRTS}. 

The prime artists explicitly provide direct access to their works but the training data artists do not. Moreover, it is unclear whether Artificial.fm has ``access" to the underlying training data because the music in the training data has been transformed into the Jukebox algorithm which does not contain copies of the works it has been trained on. Even if Artificial.fm has access to the artists' work, an infringement claim would require showing that a song created by Artificial.fm is substantially similar to a given artist's work \citep{williamsvgaye}.
Substantial similarity is assessed using a two part test: first, an objective test where a music expert analytically compares the elements of two works for substantial similarity and second, a subjective test where an ``ordinary reasonable person" assesses if the two works feel substantially similar \citep{swirskyvcarey}. 
Different experts and ``ordinary people" may disagree about substantial similarity making these tests inherently vague. In the Artificial.fm case, it is likely that some generated music is similar to some works owned by prime artists, but it is unlikely for generated music to be substantially similar to songs in the training data.

\subsection{The System Architects and Listeners \includegraphics[width=5mm]{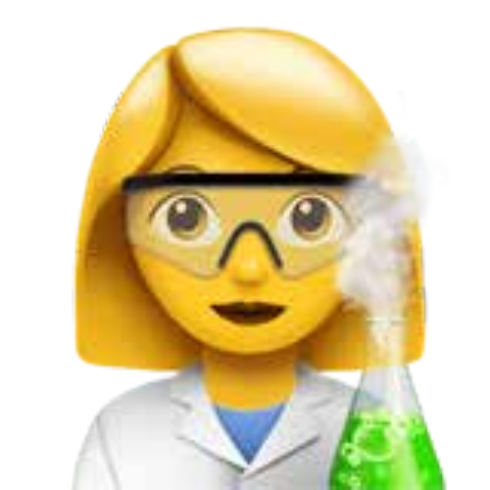} \includegraphics[width=5mm]{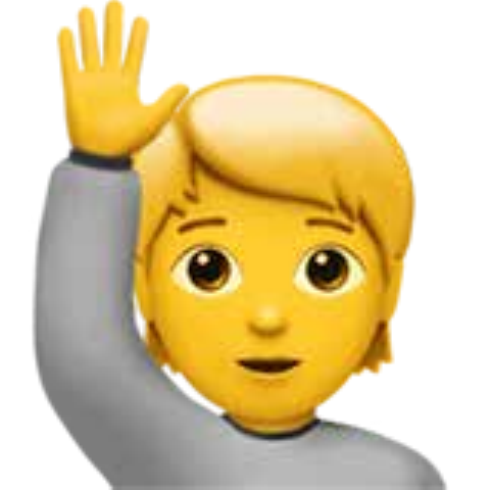}}

The system architects and listeners play their own role in creating the output of Artificial.fm and could claim ownership over the generated content. 
To focus on their contribution, imagine that Artificial.fm was trained exclusively on works in the public domain.

On one hand, the system architects might be akin to photographers who compose photographs by documenting objects from the real world.
The U.S. Supreme Court clarified in 1884 that photography is to be treated as an art under copyright law, and that the photographer is to be treated as the ``mastermind" whose creativity gives rise to a copyrightable work \citep{burrowvsarony}.
The system architects can similarly be characterized as the masterminds, who use their ingenuity to take advantage of a technology to produce works of art. On the other hand, although the listeners are using a tool built by the system architects, it is the listeners' preferences, not the architects', that guide what Artificial.fm produces. In this sense, the listeners are akin to photographers and the system architects are similar to camera makers, who have no claim to the photographs made with the technology they built.

Along these lines, the AI Artist Mario Klingemann often refers to himself as a ``neurographer,'' a photographer of neural landscapes \citep{castelle2020social}. Artificial.fm employs several design patterns so that listeners can earnestly explore the possibility space, and hence become neurographers of sorts. The personalized song recommender and preference pane push the onus of creativity onto the listener, which may in turn strengthen their ownership claim.


\subsection{The Artificial Intelligence Itself \includegraphics[width=5mm]{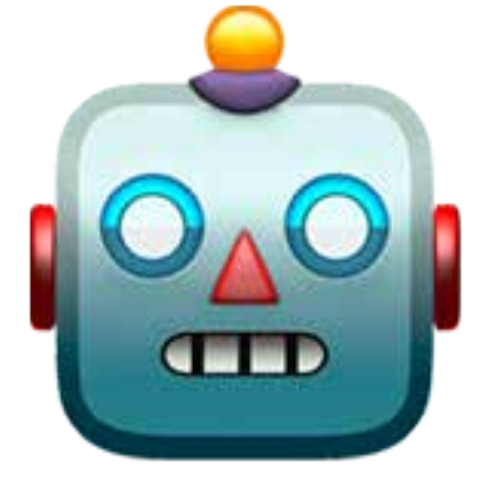}}
Perhaps the true author of Artificial.fm music is the AI \citep{compendiumcopyright}. Like a photographer, the AI decides what to create based on underlying criteria and thus identifies a small subset of expressions from a large pool of possibilities. In support of this idea, \citet{colton2020machine} present the framework of the \textit{machine condition}, by which machines creatively express their own subjectivity. However, the AI could also be compared to a sophisticated camera, a tool to enable others to create art without contributing creativity itself. Tracing the history of photography and animation, \citet{hertzmann2018can} advances this idea and argues that only \textit{social} agents can create art. \citet{epstein2020gets} find that there is natural heterogeneity in the extent to which people anthropomorphize AI (i.e. think of it as a tool vs an agent), and that these perceptions of agency are related to allocations of responsibility and credit for the involved human stakeholders.
 
If the AI \textit{is} capable of creativity, this raises the question of whether it is ``working" for whoever built it or whether it is autonomous.  In the former case, the original creator of the AI might own any creative expression created by it (under the work for hire principle \citep{bridy2012coding}). 
In the latter case, the AI might exist as some form of DAO (decentralized autonomous organization) that could be capable of ownership. 

\subsection{A Distributed Approach to Ownership}
Likely for pragmatic reasons, traditional copyright law favors resolutions with a small number of copyright owners. Many actors contribute to Artificial.fm in distinct ways, and so traditional ownership norms may be an ill fit. As a result, Artificial.fm, and platforms like it, do not fit neatly into existing ownership norms and are more suited to a distributed ownership model that divides ownership among all the actors involved in the process of casual creativity. Data cooperatives and non-fungible tokens (NFTs) are two possible technical approaches to such ownership structures. 

A data cooperative is a member-owned entity, similar to a credit union, that administers data voluntarily pooled by its members to safeguard data rights, protect privacy, and facilitate data monetization \citep{Pentland20202}. 
While data cooperatives are usually associated with personal data, they may also be useful in the context of casual creators, where all the actors who contribute to the creation of a set of works pool these works in a cooperative that advocates on behalf of all the creators.

NFTs are an application of blockchain ledgers to track the ownership of unique digital assets, which facilitates a large number of owners.  In the casual creators context, all actors involved in the creative process could receive NFTs that give them fractional ownership over one or more works. 

Both data cooperatives and NFTs are technical solutions to facilitate distributed ownership, but neither solution provides an answer to how much ownership each actor \textit{should} receive.
The normative question of how to allocate this ownership fairly and in a way that incentivizes casual creativity, is beyond the scope of this paper, but remains an open and exciting question for our community.

\section{Conclusion}



In leveraging AI for song generation, one might wonder if a formula for good music emerges. In using users' preference for songs as a metric for how good songs are, what music is perceived to be better is considerably unpredictable, making it difficult to optimize AI systems to generate ``good" music that people enjoy listening to.  Indeed, much of the time music's perceived quality is closely related with its popularity \citep{salganik2006experimental}. With music's social context being extremely influential to the public's opinion of what is good music, and gives rise to a snowball effect of ``the rich get richer,'' as the more popular songs gain more popularity while less popular songs do not see the same increase in streaming. As such, the design of the system becomes increasingly important, both to calibrate the listener's expectations for the music they will hear, and to surface cues necessary for them to make informed decisions. Casual creators like Artificial.fm bring us one step closer to understanding and integrating social context into AI systems, which in turn bootstraps their creative potential.

\section{Acknowledgements}
We would like thank Max Kreminski, Amy Smith, Simon Colton, Sandy Pentland, Abhimanyu Dubey, Christopher Bavitz, Jessica Fjeld, Daza Greenwood, Matt Groh and Jack Muller for helpful feedback and discussion. We would also like the thank the musicians who provided primes to Artificial.fm, and the listeners of Artificial.fm, as well as the artists whose music Jukebox was trained on.

\section*{Author Contributions}
SG, MH, MM and ZE conceptualized the project. SG and ZE developed the system. SG, MH, MM and ZE wrote the paper. 

\bibliographystyle{iccc}
\bibliography{iccc}

\begin{thebibliography}{}

\bibitem[\protect\citeauthoryear{{ Burrow-Giles Lithographic Co. v.
  Sarony}}{1884}]{burrowvsarony}
{ Burrow-Giles Lithographic Co. v. Sarony}.
\newblock 1884.
\newblock Supreme court of the united states.

\bibitem[\protect\citeauthoryear{Blaauw and Bonada}{2017}]{blaauw2017neural}
Blaauw, M., and Bonada, J.
\newblock 2017.
\newblock A neural parametric singing synthesizer.
\newblock {\em arXiv preprint arXiv:1704.03809}.

\bibitem[\protect\citeauthoryear{Bogdanov and
  others}{2013}]{bogdanov2013essentia}
Bogdanov, D., et~al.
\newblock 2013.
\newblock Essentia: An audio analysis library for music information retrieval.
\newblock In {\em 14th Conference of the International Society for Music
  Information Retrieval}.
\newblock ISMIR.

\bibitem[\protect\citeauthoryear{Bonada and Serra}{2007}]{bonada2007synthesis}
Bonada, J., and Serra, X.
\newblock 2007.
\newblock Synthesis of the singing voice by performance sampling and spectral
  models.
\newblock {\em IEEE signal processing magazine} 24(2):67--79.

\bibitem[\protect\citeauthoryear{Bridy}{2012}]{bridy2012coding}
Bridy, A.
\newblock 2012.
\newblock Coding creativity: copyright and the artificially intelligent author.
\newblock {\em Stan. Tech. L. Rev.} ~5.

\bibitem[\protect\citeauthoryear{Castelle}{2020}]{castelle2020social}
Castelle, M.
\newblock 2020.
\newblock The social lives of generative adversarial networks.
\newblock In {\em FAT*},  413.

\bibitem[\protect\citeauthoryear{Chapelle and Li}{2011}]{chapelle2011empirical}
Chapelle, O., and Li, L.
\newblock 2011.
\newblock An empirical evaluation of thompson sampling.
\newblock {\em Advances in neural information processing systems}
  24:2249--2257.

\bibitem[\protect\citeauthoryear{Colton \bgroup et al.\egroup
  }{2020}]{colton2020machine}
Colton, S.; Pease, A.; Guckelsberger, C.; McCormack, J.; Llano, T.; et~al.
\newblock 2020.
\newblock On the machine condition and its creative expression.
\newblock In {\em International Conference on Computational Creativity}.

\bibitem[\protect\citeauthoryear{Compton and Mateas}{2015}]{compton2015casual}
Compton, K., and Mateas, M.
\newblock 2015.
\newblock Casual creators.
\newblock In {\em ICCC},  228--235.

\bibitem[\protect\citeauthoryear{Compton}{2019}]{compton2019casual}
Compton, K.
\newblock 2019.
\newblock {\em Casual creators: Defining a genre of autotelic creativity
  support systems}.
\newblock University of California, Santa Cruz.

\bibitem[\protect\citeauthoryear{Dhariwal \bgroup et al.\egroup
  }{2020}]{dhariwal2020jukebox}
Dhariwal, P.; Jun, H.; Payne, C.; Kim, J.~W.; Radford, A.; and Sutskever, I.
\newblock 2020.
\newblock Jukebox: A generative model for music.
\newblock {\em arXiv preprint arXiv:2005.00341}.

\bibitem[\protect\citeauthoryear{Draves}{2005}]{draves2005electric}
Draves, S.
\newblock 2005.
\newblock The electric sheep screen-saver: A case study in aesthetic evolution.
\newblock In {\em Workshops on Applications of Evolutionary Computation},
  458--467.
\newblock Springer.

\bibitem[\protect\citeauthoryear{Engel \bgroup et al.\egroup
  }{2017}]{engel2017neural}
Engel, J.; Resnick, C.; Roberts, A.; Dieleman, S.; Norouzi, M.; Eck, D.; and
  Simonyan, K.
\newblock 2017.
\newblock Neural audio synthesis of musical notes with wavenet autoencoders.
\newblock In {\em International Conference on Machine Learning},  1068--1077.
\newblock PMLR.

\bibitem[\protect\citeauthoryear{Epstein \bgroup et al.\egroup
  }{2020a}]{epstein2020interpolating}
Epstein, Z.; Boulais, O.; Gordon, S.; and Groh, M.
\newblock 2020a.
\newblock Interpolating gans to scaffold autotelic creativity.
\newblock {\em arXiv preprint arXiv:2007.11119}.

\bibitem[\protect\citeauthoryear{Epstein \bgroup et al.\egroup
  }{2020b}]{epstein2020gets}
Epstein, Z.; Levine, S.; Rand, D.~G.; and Rahwan, I.
\newblock 2020b.
\newblock Who gets credit for ai-generated art?
\newblock {\em Iscience} 23(9):101515.

\bibitem[\protect\citeauthoryear{Epstein \bgroup et al.\egroup
  }{2021}]{epstein2021social}
Epstein, Z.; Groh, M.; Dubey, A.; and Pentland, A.
\newblock 2021.
\newblock Social influence leads to the formation of diverse local trends.
\newblock {\em Proceedings of the ACM on Human-Computer Interaction}
  5(CSCW2):1--18.

\bibitem[\protect\citeauthoryear{{Feist v. Rural Telephone Service
  Co.}}{1991}]{FeistvRTS}
{Feist v. Rural Telephone Service Co.}
\newblock 1991.
\newblock Supreme court of the united states.

\bibitem[\protect\citeauthoryear{Hadjeres, Pachet, and
  Nielsen}{2017}]{hadjeres2017deepbach}
Hadjeres, G.; Pachet, F.; and Nielsen, F.
\newblock 2017.
\newblock Deepbach: a steerable model for bach chorales generation.
\newblock In {\em ICML},  1362--1371.
\newblock PMLR.

\bibitem[\protect\citeauthoryear{Hertzmann}{2018}]{hertzmann2018can}
Hertzmann, A.
\newblock 2018.
\newblock Can computers create art?
\newblock In {\em Arts}, volume~7, ~18.
\newblock Multidisciplinary Digital Publishing Institute.

\bibitem[\protect\citeauthoryear{Hiller~Jr and
  Isaacson}{1957}]{hiller1957musical}
Hiller~Jr, L.~A., and Isaacson, L.~M.
\newblock 1957.
\newblock Musical composition with a high speed digital computer.
\newblock In {\em Audio Engineering Society Convention 9}.
\newblock Audio Engineering Society.

\bibitem[\protect\citeauthoryear{Huang \bgroup et al.\egroup
  }{2018}]{huang2018music}
Huang, C.-Z.~A.; Vaswani, A.; Uszkoreit, J.; Shazeer, N.; Hawthorne, C.; Dai,
  A.~M.; Hoffman, M.~D.; and Eck, D.
\newblock 2018.
\newblock Music transformer: Generating music with long-term structure.
\newblock {\em arXiv preprint arXiv:1809.04281}.

\bibitem[\protect\citeauthoryear{Lewis}{2000}]{lewis2000too}
Lewis, G.~E.
\newblock 2000.
\newblock Too many notes: Computers, complexity and culture in voyager.
\newblock {\em Leonardo Music Journal}  33--39.

\bibitem[\protect\citeauthoryear{Maconie}{1970}]{stockhausen}
Maconie, R.
\newblock 1970.
\newblock Stockhausen's 'setz die segel zur sonne'.
\newblock {\em Tempo} (92):30--32.

\bibitem[\protect\citeauthoryear{Mehri \bgroup et al.\egroup
  }{2016}]{mehri2016samplernn}
Mehri, S.; Kumar, K.; Gulrajani, I.; Kumar, R.; Jain, S.; Sotelo, J.;
  Courville, A.; and Bengio, Y.
\newblock 2016.
\newblock Samplernn: An unconditional end-to-end neural audio generation model.
\newblock {\em arXiv preprint arXiv:1612.07837}.

\bibitem[\protect\citeauthoryear{Moorer}{1972}]{moorer1972music}
Moorer, J.~A.
\newblock 1972.
\newblock Music and computer composition.
\newblock {\em Communications of the ACM} 15(2):104--113.

\bibitem[\protect\citeauthoryear{Pentland and Hardjono}{2020}]{Pentland20202}
Pentland, A., and Hardjono, T.
\newblock 2020.
\newblock 2. data cooperatives.
\newblock In {\em Building the New Economy}. 0 edition.
\newblock https://wip.mitpress.mit.edu/pub/pnxgvubq.

\bibitem[\protect\citeauthoryear{Rappaz \bgroup et al.\egroup
  }{2018}]{rappaz2018latent}
Rappaz, J.; Catasta, M.; West, R.; and Aberer, K.
\newblock 2018.
\newblock Latent structure in collaboration: the case of reddit r/place.
\newblock In {\em Twelfth International AAAI Conference on Web and Social
  Media}.

\bibitem[\protect\citeauthoryear{Salganik, Dodds, and
  Watts}{2006}]{salganik2006experimental}
Salganik, M.~J.; Dodds, P.~S.; and Watts, D.~J.
\newblock 2006.
\newblock Experimental study of inequality and unpredictability in an
  artificial cultural market.
\newblock {\em Science} 311(5762):854--856.

\bibitem[\protect\citeauthoryear{Secretan \bgroup et al.\egroup
  }{2011}]{secretan2011picbreeder}
Secretan, J.; Beato, N.; D'Ambrosio, D.~B.; Rodriguez, A.; Campbell, A.;
  Folsom-Kovarik, J.~T.; and Stanley, K.~O.
\newblock 2011.
\newblock Picbreeder: A case study in collaborative evolutionary exploration of
  design space.
\newblock {\em Evolutionary computation} 19(3):373--403.

\bibitem[\protect\citeauthoryear{{Swirsky v. Carey}}{2004}]{swirskyvcarey}
{Swirsky v. Carey}.
\newblock 2004.
\newblock United states court of appeals for the ninth circuit.

\bibitem[\protect\citeauthoryear{{United States Copyright
  Office}}{2021}]{compendiumcopyright}
{United States Copyright Office}.
\newblock 2021.
\newblock Compendium of us copyright office practices.

\bibitem[\protect\citeauthoryear{{Williams v. Gaye}}{2018}]{williamsvgaye}
{Williams v. Gaye}.
\newblock 2018.
\newblock United states court of appeals for the ninth circuit.

\bibitem[\protect\citeauthoryear{Xenakis}{1992}]{xenakis1992formalized}
Xenakis, I.
\newblock 1992.
\newblock {\em Formalized music: thought and mathematics in composition}.
\newblock Pendragon Press.

\bibitem[\protect\citeauthoryear{Yannakakis, Liapis, and
  Alexopoulos}{2014}]{yannakakis2014mixed}
Yannakakis, G.~N.; Liapis, A.; and Alexopoulos, C.
\newblock 2014.
\newblock Mixed-initiative co-creativity.

\end{thebibliography}


\begin{thebibliography}{4}
\providecommand{\natexlab}[1]{#1}
\providecommand{\url}[1]{\texttt{#1}}
\expandafter\ifx\csname urlstyle\endcsname\relax
  \providecommand{\doi}[1]{doi: #1}\else
  \providecommand{\doi}{doi: \begingroup \urlstyle{rm}\Url}\fi

\bibitem[Dhariwal et~al.(2020)Dhariwal, Jun, Payne, Kim, Radford, and
  Sutskever]{dhariwal2020jukebox}
Prafulla Dhariwal, Heewoo Jun, Christine Payne, Jong~Wook Kim, Alec Radford,
  and Ilya Sutskever.
\newblock Jukebox: A generative model for music.
\newblock \emph{arXiv preprint arXiv:2005.00341}, 2020.

\bibitem[Bogdanov et~al.(2013)]{bogdanov2013essentia}
Dmitry Bogdanov et~al.
\newblock Essentia: An audio analysis library for music information retrieval.
\newblock In \emph{14th Conference of the International Society for Music
  Information Retrieval}. ISMIR, 2013.

\bibitem[Thompson(1933)]{thompson1933likelihood}
William~R Thompson.
\newblock On the likelihood that one unknown probability exceeds another in
  view of the evidence of two samples.
\newblock \emph{Biometrika}, 25\penalty0 (3/4):\penalty0 285--294, 1933.

\bibitem[Chapelle and Li(2011)]{chapelle2011empirical}
Olivier Chapelle and Lihong Li.
\newblock An empirical evaluation of thompson sampling.
\newblock \emph{Advances in neural information processing systems},
  24:\penalty0 2249--2257, 2011.

\end{thebibliography}


\end{document}


\maketitle
\beginsupplement

\section{System Overview}

\begin{figure*}[h]
    \centering
    \includegraphics[width=\textwidth]{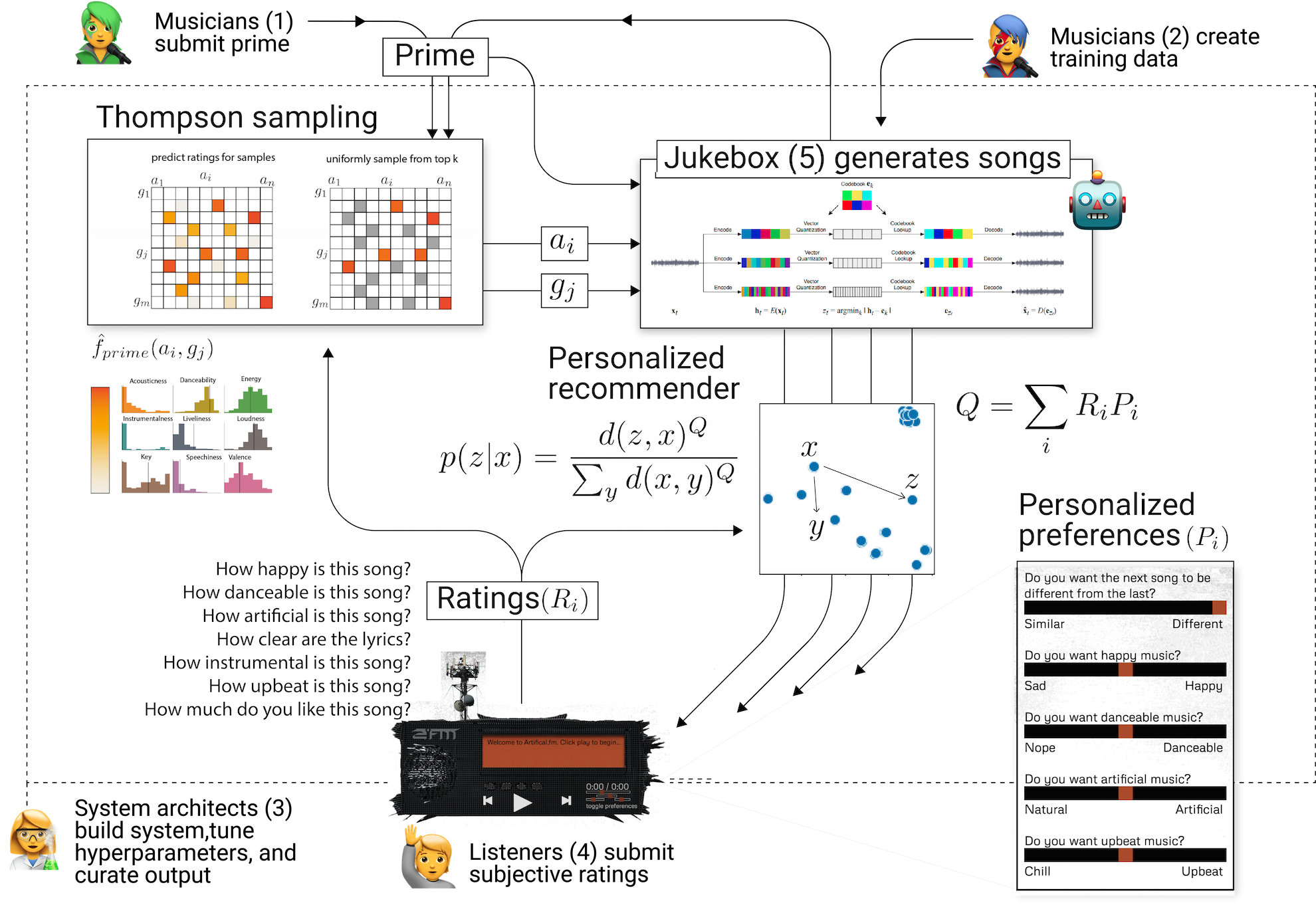}
    \caption{System map of the Artificial.fm platform. Explore live at \url{www.artificial.fm}  }
    \label{fig:map}
\end{figure*}

A system map for Artificial.fm is shown in Figure~\ref{fig:map}. The interface is built with Howler.js and p5.js. 
\subsection{Song Generation with Jukebox}
We forked the Jukebox repository \cite{dhariwal2020jukebox} to build our song generation pipeline. All experiments were run with a NVIDIA GeForce GTX TITAN X. The songs were generated with genre prompts from folk, house, pop, americana, rock, classical, electronic, and funk, and artist prompts from The Weeknd, Aerosmith, The Doors, Justin Bieber, Elton John, Dolly Parton, Otis Redding, and Lady Gaga. The primes were sourced from several local artists we reached out to.

\subsection{Crowdsourced Subjective Evaluation}
\begin{figure*}[h]
    \centering
    \includegraphics[width=\textwidth]{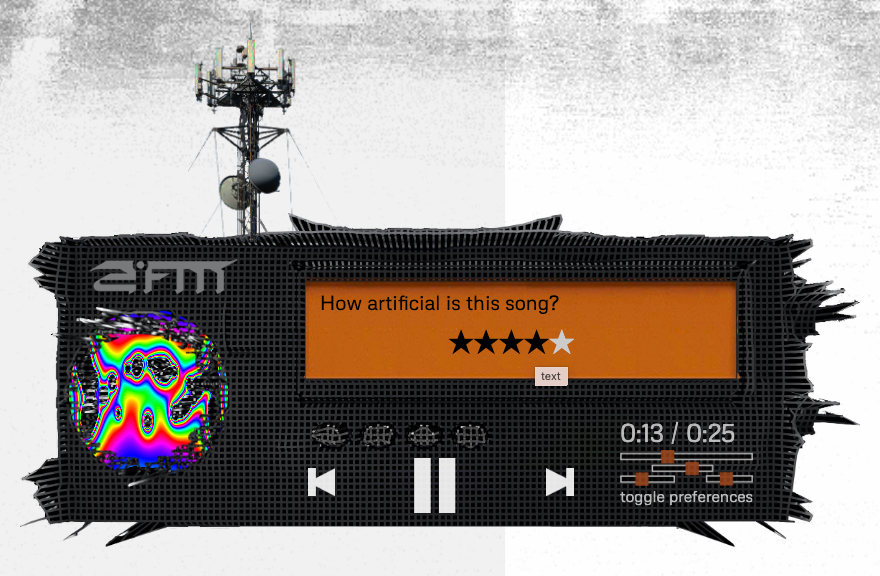}
    \caption{Rating interface for Artificial.fm }
    \label{fig:rating}
\end{figure*}

The outputs of the song generation process are streamed via the platform, where listeners provide subjective feedback on the AI-generated outputs, in the form of ratings. The questions asked about each song were in the form of a 5-point likert scale with stars (shown in Figure~\ref{fig:rating}), and the seven questions are listed in Table~\ref{fig:map}. These questions are asked one by one in a new random order for each song.  The collection of this data was exempted by MIT COUHES.

\begin{table*}[h]
\centering
  \caption{Questions for subjective rating}
  \label{tab:share_polit}
\begin{tabular}{l}
  \hline
 How happy is this song? \\ 
 How danceable is this song? \\ 
 How artificial is this song? \\ 
 How clear are the lyrics? \\ 
 How instrumental is this song? \\ 
 How upbeat is this song? \\ 
 How much do you like this song? \\ 

   \hline
\end{tabular}
\end{table*}

\subsection{Covariates for song generation}
We use the Spotify API and Essentia \cite{bogdanov2013essentia} to generate a rich set of covariates for the artist of the prime, as well as candidate artists and genres. For an artist, we find the 10 most popular songs for that artist, and compute the average Essentia features \citep{bogdanov2013essentia} of those songs (danceability, energy, key, loudness, speechiness, acousticness, instrumentalness, liveness, valence). For a given genre, we find the top 20 most popular Spotify playlists associated with that genre, and then compute the average Essentia features for all songs in those playlists.

\subsection{Generating new songs}
The crowdsourced feedback is then used to adapt the generation process with an algorithm that balances exploring new permutations of parameters with exploiting existing parameters that are popular with users. This is achieved using a variation of Thompson sampling \citep{thompson1933likelihood}, which is regret-minimizing in such contexts \citep{chapelle2011empirical}. As new primes are solicited from local musicians, the following algorithm finds parameters (e.g. an artist and genre prompt) to pair with that prime to balance exploration and exploitation (see Algorithm~\ref{alg:thompson}):  First, we fit a model $\hat{f}$ predicting ratings of the existing songs based on the Spotify covariates of that song's prime artist, artist prompt, and genre prompt (27 features total). Then, we sample M artist, genre pairs from the joint distribution of these prompts in the input space. Then, we predict the rating of that artist, genre pair for the given prompt $\hat{f}_{prime}(a_\ell, g_\ell)$. We then take the top $\gamma$ artist, genre pairs and randomly sample one uniformly (here $\gamma$ controls exploration vs exploitation, $\gamma=1$ is maximal exploitation, $\gamma=M$ is maximal exploration).

A final missing ingredient is the actual outcome variable to optimize. The most straight forward variable is response ratings to the ``How much do you like this song?'' question. But we can explore other outcomes to optimize, such as the variation in how much people like the song (e.g. the song's contraversional-ness), or some weighted average of all seven features, based on the personalized preferences listeners input (see section below). 

\begin{algorithm}[H]
  \KwIn{prime, $\gamma$, M}
  \KwOut{k}
  $\hat{y}_{prime} = f_{prime}(a_i, g_j)$\; 
  \For{$k = 1$ to $M$}{
    $ (a_k, g_k) \sim P(a,g)$\;
    $\hat{y}_k = f_{prime}(a_k, g_k)$\;
  }
  $F = \text{sort}(\{\hat{y}_k\})$ \;
  $\ell \sim F[1...\gamma]$\;
  \Return $\ell$
  \caption{Thompson Sampling}
  \label{alg:thompson}
\end{algorithm}

\subsection{Personalized Song Recommendations}
\begin{figure*}[h]
    \centering
    \includegraphics[width=\textwidth]{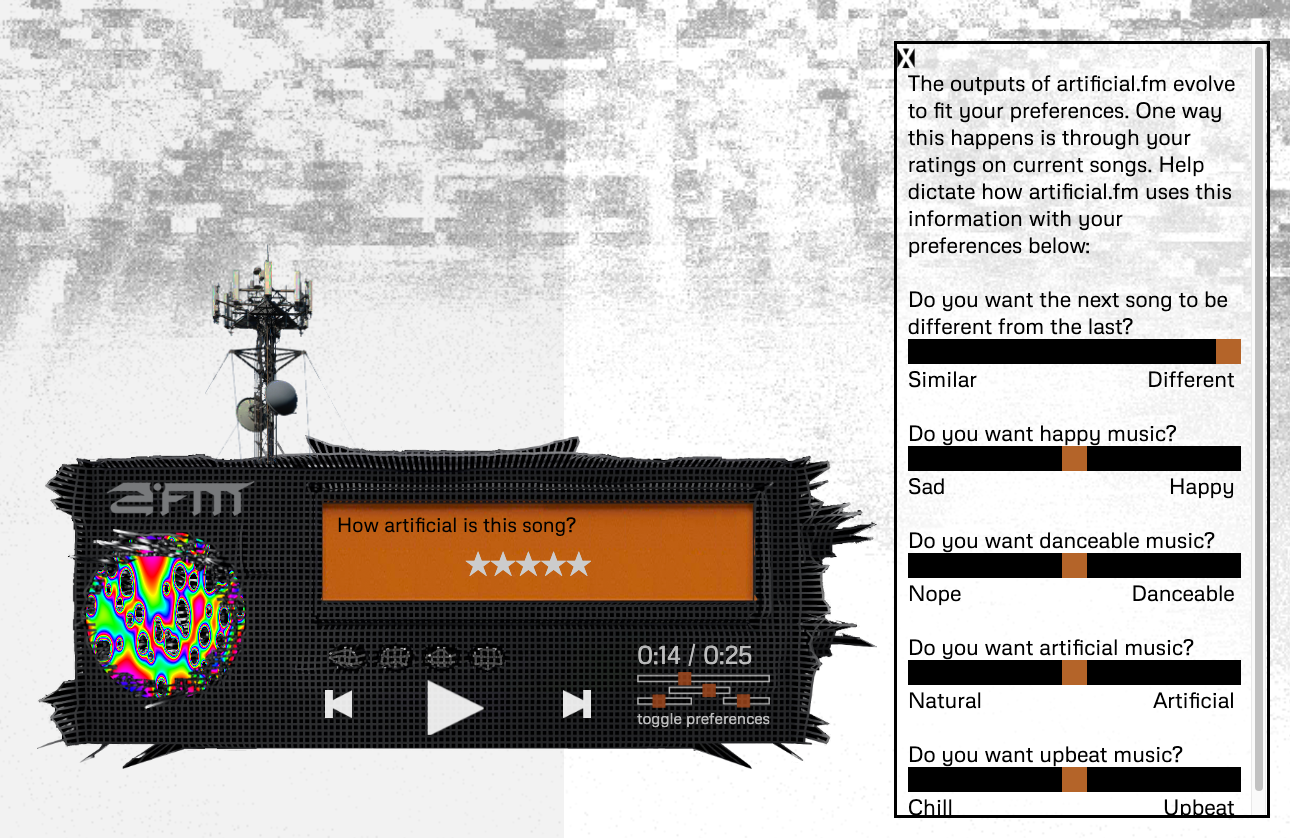}
    \caption{User preference elicitation interface. }
    \label{fig:guide}
\end{figure*}

The order of the songs streamed to a listener is determined by a personalized song recommender, based on both their ratings of songs, and stated preferences. Listener's can express their preferences $P_i$ through the ``personalized preferences'' panel found on the main radio interface (see Figure~\ref{fig:guide}). This panel allows listeners to assign a numerical weight in the range [-2,2] to five aspects of songs: difference from current song (initialized to 2), happiness (initialized to 0), danceability (initialized to 0), artificiality (initialized to 0) and upbeat-ness (initialized to 0). This personalized preferences panel gives listeners more agency by giving them a grammar to navigate through the possibility space. 

When a song $x$ ends and it is time to determine the next song $z$ to be played, a \textit{quality score} for song $x$ is computed as the sum of ratings, weighted by personalized preference scores:
$$Q_x = \sum_i R_iP_i$$
Since both $R_i$ and $P_i$ are bounded by [-2,2], a positive $Q$ means the listener wants a song different to the previous, while a negative $Q$ means they want something similar. Thus, given $d(x,y)$ as the euclidean distance between songs using the Spotify covariates, we sample the next song to play, $z$, with probability

$$p(z|x) = \frac{d(z,x)^{Q_x/B}}{\sum_y d(x, y)^{Q_x/B}}$$

where $y$ are the songs that have not yet been played and $B$ is a tuning parameter. $Q_x<0$ means a new song is sampled proportional to its similarity with the previous song. $Q_x>0$ means a new song is sampled inversely proportional to its similarity with the previous song, while $Q_x =0$ means songs are sampled uniformly.


\bibliography{iccc}